# Automatic Detection of Indoor and Outdoor Scenarios using NMEA Message Data from GPS Receivers

R.S. Pissardini and E.S. Fonseca Junior

*Abstract*—Detection of indoor and outdoor scenarios is an important resource for many types of activities such as multisensor navigation and location-based services. This research presents the use of NMEA data provided by GPS receivers to characterize different types of scenarios automatically. A set of static tests was performed to evaluate metrics such as number of satellites, positioning solution geometry and carrier-to-receiver noise-density ratio values to detect possible patterns to determine indoor and outdoor scenarios. Subsequently, validation tests are applied to verify that parameters obtained are adequate.

## I. Introdução

GNSSes (plural do inglês *Global Navigation Satellite System* - sistema de navegação global por satélite) é o nome genérico dado aos sistemas de posicionamento e navegação (POS/NAV) capazes de prover solução de posição tridimensional a partir de uma constelação de satélites dedicada a transmitir dados posicionais sobre a superfície terrestre. Com a aquisição do sinal de um mínimo de quatro satélites por parte de um receptor apropriado, as coordenadas da posição podem ser calculadas através de trilateração. Exemplos de sistemas GNSS operacionais incluem o estadunidense GPS (em inglês *Global Positioning System*), o russo GLONASS (em russo *Globalnaya Navigatsionnaya Sputnikovaya Sistema*) e o europeu GALILEO, que podem ser utilizados de forma independente ou integrada em soluções multiconstelação [1]–[4].

Idealmente, o principal cenário de operação para um receptor GNSS é o ambiente externo e sem obstruções, isto é, o ambiente onde possíveis bloqueios naturais (elevações de terreno, vegetação, etc.) e artificiais (edificações, obras de infraestrutura, etc.) que impeçam a visada direta entre os centros de fase da antena do satélite de uma determinada constelação e da antena do receptor do usuário não existam ou sejam irrelevantes [5]. Para atividades tradicionais de Geomática como levantamento topográfico, receptores GNSS são configurados e posicionados na melhor situação possível que permita a aquisição da maior quantidade de sinais possíveis em visada direta.

No entanto, em outras necessidades de dados de POS/NAV, os ambientes podem oferecer restrições que inviabilizam o uso exclusivo de tecnologias GNSS. Exemplos incluem posicionamento para atividades de transportes rodoviários, para serviços baseados em localização (em inglês *Location-Based Services* - LBS), entre outros. Em [5], os autores sistematizam as principais propostas da literatura para minimizar os efeitos de bloqueios aos GNSSes. As principais técnicas neste sentido podem ser agrupadas em dois grandes grupos: (a) aquelas que buscam melhorar a aquisição de sinais GNSS em ambientes de dificuldade a este tipo de tecnologia e (b) aquelas que utilizam outras tecnologias de navegação (e.g. sistemas inerciais) como substitutas ou como auxílio às tecnologias GNSS quando em ambientes de dificuldade.

Independente da abordagem utilizada, a detecção automática de cenários internos e externos por parte de um sistema de navegação é recomendada para permitir que tal sistema possa definir, dentre os recursos disponíveis, as tecnologias mais apropriadas para operação (em especial, se o sistema deve transitar entre cenários heterogêneos). Isto é particularmente importante para sistemas de navegação multisensor, tecnologias de consciência de contexto e para LBS. A literatura científica apresenta diversos tipos de propostas para suportar este tipo de detecção automática, em especial, para sistemas baseados em contexto em tecnologias móveis: exemplos incluem uso de câmeras de vídeo, potência de sinal de telefonia celular, sinais de redes sem fio, iluminação do ambiente, intensidade magnética e outros[6]–[10].

No presente artigo avalia-se a utilização de dados obtidos de mensagens NMEA (do inglês *National Marine Electronics Association)* de um receptor GPS para realização deste tipo de análise de forma simplificada: em geral todos os receptores GNSS, independentemente da sua qualidade e dos recursos fornecidos, oferecem este conjunto de mensagens ao contrário de outros tipos de dados que são dependentes dos acessos fornecidos por fabricantes específicos.

R.S. Pissardini, Laboratório de Topografia e Geodesia, Departamento de Engenharia de Transportes, Escola Politécnica da Universidade de São Paulo (EPUSP), São Paulo, Brazil (e-mail: pissardini@usp.br).

E.S. Fonseca Junior, Laboratório de Topografia e Geodesia, Departamento de Engenharia de Transportes, Escola Politécnica da Universidade de São Paulo (EPUSP), São Paulo, Brazil (e-mail: edvaldoj@usp.br).

## II. Detecção de cenários de operação utilizando sinais GPS

A definição e distinção entre cenário interno (em inglês *indoor*) e cenário externo (em inglês *outdoor*) não possui uma definição consensual na literatura científica [5]. Por convenção, um cenário interno é aquele que pertence ao interior de uma edificação (e no qual sinais GNSS sofrem maior bloqueio ou atenuação), enquanto um cenário externo é aquele que não está limitado por uma edificação (e por isto sinais GNSS sofreriam menor quantidade de bloqueios e atenuação). Esta definição, no entanto, pode não ser adequada em diversos casos já que há cenários internos cujo comportamento é semelhante ao de cenários externos e vice-versa. Este artigo não traz uma discussão sobre a definição formal do que é um cenário interno ou cenário externo, mas se é possível associar características obtidas de receptores GNSSes a cenários que seres humanos comumente identificam como internos ou externos.

Intuitivamente, uma hipótese ingênua pode estimar que receptor GNSS está em um ambiente interno ou em um ambiente externo pela quantidade de sinais de satélites recebidos. Se um receptor consegue obter quatro ou mais satélites para calcular sua posição, pode considerar-se que o receptor está em um cenário externo aberto sem bloqueios relevantes, enquanto a situação oposta caracteriza um cenário interno [6]. No entanto, esta abordagem pode se mostrar inadequada pois dependendo dos materiais de construção utilizados para estruturar uma determinada edificação, os cenários internos associados podem permitir que um receptor obtenha sinais suficientes para determinar sua posição (mesmo que estes sinais possam chegar atenuados, degradados ou sem visada direta), enquanto alguns cenários externos podem apresentar severos bloqueios de sinais, em especial, em situações de vias estreitas e cânions urbanos [5]. Nesta pesquisa, avalia-se se a quantidade de satélites pode ser utilizada, de alguma forma, para caracterizar ambientes internos. Outros dois parâmetros oferecidos pelas mensagens NMEA de receptores GPS são também avaliados com este propósito: (a) a geometria dos satélites em relação ao receptor e (b) a análise dos valores da relação portadora-densidade do ruído para os satélites disponíveis.

Diluição da Precisão (em inglês *Dilution of Precision*- DOP) é o nome da representação numérica da geometria entre os satélites e o receptor em um determinado instante de tempo, sendo que seu valor é associado ao erro posicional [1][4][6]. A disposição ideal é que os satélites estejam igualmente espaçados, pois se os satélites estão próximos entre si ou não se espalham de forma uniforme no horizonte de observação, a sobreposição entre os sinais pode aumentar o erro da posição calculada e limitar a qualidade da solução de navegação. Matematicamente, DOP representa a razão entre os desvios-padrão de um dado parâmetro em relação às pseudodistâncias, sendo que as equações relacionadas podem ser encontradas em [11]. Os valores adequados de DOP, assim, devem ser menores do que 4, enquanto valores maiores do que 10 representam que a solução não deveria ser utilizada.

Receptores GNSS calculam vários tipos de DOP: Diluição Vertical de Precisão (VDOP – 1-D), Diluição Horizontal de Precisão (HDOP – 2-D), Diluição da Precisão da Posição (PDOP – 3-D) (combinação entre HDOP e VDOP) e Diluição da Precisão do Tempo (TDOP). A combinação entre PDOP e TDOP fornece a Diluição Geométrica da Posição (GDOP). Um outro tipo de medição chamada de Diluição da Precisão Relativa (RDOP) pode ser calculado quando se realiza posicionamento relativo [12], [13]. Em relação à detecção de cenários internos e externos, pelas suas características, o DOP de uma dada solução tende a piorar se o receptor está em uma região de bloqueios que inviabiliza a detecção dos satélites de forma adequada. Enquanto isto pode ser associado com o fato de que o receptor está em uma região interna, deve-se considerar que lugares abertos externos podem apresentar valores de DOP inadequados devido à distribuição de bloqueios no horizonte de observação.

A medição da relação portadora-densidade do ruído ($C/N_0$) está relacionada com as características da propagação de ondas eletromagnéticas dos sinais GNSS. $C/N_0$ expressa a razão (em dB-Hz) entre a potência da onda portadora e a potência do ruído por unidade de largura de banda. O valor teórico de $C/N_0$ em dB-Hz é dado pela fórmula [9][10]:

$$\frac{C}{N_0} = S_r + G_a - 10 \times log_{10}(k) - 1 \times log_{10}(T_{sys}) - L \quad (1)$$

Na qual:
- $S_r$: é a potência do sinal de um dado satélite.
- $G_a$: é o ganho da antena em relação ao satélite.
- $k$: é a constante de Boltzman ($1.38 \times 10^{-23}\ Watt - sec/K$).
- $T_{sys}$: é a temperatura do ruído do sistema, dada por ($T_{source} + T_{receiver}$) onde $T_{source}$ é o ruído de temperatura na fonte do sinal e $T_{receiver}$ no receptor.
- $L$: perdas de implementação.

$C/N_0$ não deve ser confundida com a medição *Signal-to-Noise Ratio* (SNR) que é, também, outra maneira que alguns receptores GNSS expressam a potência com a qual um receptor recebe o sinal de um satélite. SNR é a relação entre a potência do sinal e a potência do ruído em uma determinada largura de banda [14]. Sua fórmula, expressa em dB, é dada por:

$$SNR = S_r - N \quad (2)$$

Na qual:
- $S_r$: é a potência do sinal de um dado satélite.
- $N$: é a potência do ruído.

Uma relação entre $C/N_0$ e SNR pode ser estabelecida sob a forma [14]:

$$C/N_0 = SNR + BW \quad (3)$$

Onde $BW$ é a largura de banda da observação.

Idealmente, um sinal GNSS deve ser obtido em visada direta (em inglês *line-of-sight* – LOS) entre o satélite GNSS e o receptor do usuário (i.e., entre os centros de fase de suas antenas), sem interferências de bloqueio de outros elementos na

Oct/2017

propagação. No entanto, como os sinais são propagados através de diversos tipos de materiais (incluindo as camadas da atmosfera) o sinal recebido por um receptor é, em geral, bastante atenuado [5]. $C/N_0$ é, neste contexto, um indicador mais recomendado para mensurar o nível de atenuação e a qualidade com a qual o sinal de um satélite é recebido pela antena do receptor GNSS, de forma independente aos algoritmos internos deste receptor do que SNR. Dado este pressuposto, pode-se considerar que regiões externas e abertas devem possuir menor atenuação da potência do sinal do que regiões com bloqueios mais densos e que cenários internos em regiões urbanas devem apresentar comportamento semelhante devido aos tipos de materiais utilizados para sua construção (Tabela I).

TABELA I
ATENUAÇÕES APRESENTADAS POR ALGUNS MATERIAIS DE CONSTRUÇÃO PARA SINAIS NA BANDA-L [16]

| Material | Atenuação (dB) | Fator (-) |
| --- | --- | --- |
| *Drywall* | 1 | 0.8 |
| Madeira compensada | 1-3 | 0.8-0.5 |
| Vidro | 1-4 | 0.8 -0.4 |
| Madeira | 2-9 | 0.6-0.1 |
| Grade de vergalhão | 2-11 | 0.6-0.08 |
| Tijolo | 5-31 | 0.3-0.001 |
| Concreto | 12-43 | 0.06-0.00005 |
| Concreto armado | 29-33 | 0.001-0.0005 |

A primeira proposta para utilização de valores de $C/N_0$ para determinar se um receptor está em um ambiente interno ou em um ambiente externo é descrita em [17], que também cita a utilização de estimadores baseados em Fator Riciano (K), servindo como base para a utilização rudimentar em uma técnica descrita em [7]. Em [7] utiliza-se um *smartphone* para obter dados NMEA e a medição de $C/N_0$ de sinais GNSS em um determinado ambiente. Sobre os valores de $C/N_0$ para uma determinada época de medição são calculadas a soma, a média e os respectivos valores de desvios-padrões. Os testes foram realizados em 15 diferentes localidades com aquisição de 100 segundos de dados (com uma frequência de 1 Hz) em cada local. Os resultados demonstraram que ambientes tidos como internos possuem uma média de $C/N_0$ menor que 25 dB-Hz com baixo desvio-padrão, enquanto ambientes externos possuem uma média de $C/N_0$ maior que 30 dB-Hz com desvio-padrão maior. Os resultados também apresentaram que a soma dos valores de $C/N_0$ para um mesmo instante deve ser maior do que 200 dB-Hz em ambientes externos, enquanto ambientes internos devem possuir um valor inferior a este limite. Deve-se considerar, no entanto, que *smartphones* possuem restrições e filtros próprios que podem não ser de fácil manipulação e desligamento de modo que para testes mais confiáveis recomenda-se a utilização de receptores mais adequados. Uma continuação dos testes foram apresentadas pelos autores em [18].

Oct/2017

III. METODOLOGIA, TESTES E RESULTADOS

*A. Padrões de comportamento de cenários internos e externos*

Na primeira fase de testes, esta pesquisa avalia a utilização das métricas de quantidades de satélites, média e soma dos valores de $C/N_0$ e os valores de DOP para determinar se um ambiente é interno ou externo. Busca-se, assim, detectar os possíveis padrões associados a uma métrica específica ou em combinação com outras que permita realizar este tipo de caracterização.

Nesta etapa, foi realizado um conjunto de testes estáticos utilizando coletores GNSS combinando antenas de uso geral, receptores GNSS U-Blox Lea 6T (L1 apenas com GPS, com todos os possíveis filtros - e.g., limites de número de satélites, valor de geometria e ruído do sinal – desabilitados)), minicomputadores Raspberry Pi para armazenamento e processamento de dados e baterias para garantir a mobilidade do equipamento (Fig. 1). Todos os coletores utilizados foram igualmente configurados e calibrados de forma a oferecer processamento semelhante operando sob as mesmas condições. Foi desenvolvido um *software* embarcado para aquisição de dados NMEA com a frequência de 1 Hz.

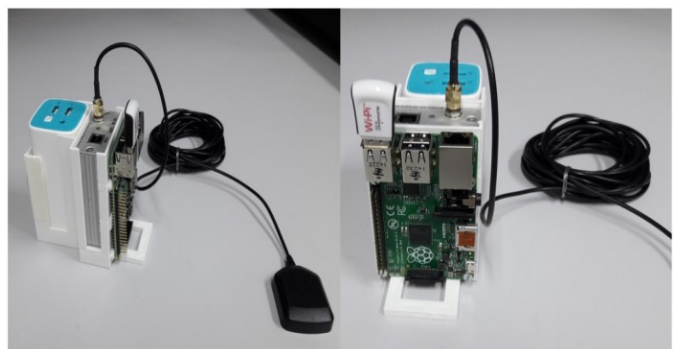

Fig. 1. Receptores GNSS U-Blox Lea 6T integrados em um minicomputador Raspberry Pi com baterias.

Para os testes estáticos foram definidos quatro tipos de cenários para serem analisados (Tabela II). Estes cenários foram definidos convencionalmente, optando-se por não estabelecer uma divisão direta entre cenário interno e cenário externo já que, no limite entre estes dois tipos de cenários, há sub-cenários que podem necessitar de caracterizações mais adequadas. Esta divisão possui paralelos com a distinção apresentada em [6] e [18].

TABELA II
TIPOS DE CENÁRIOS E CARACTERÍSTICAS CONVENCIONADAS

| ID | Local | Características |
| --- | --- | --- |
| 1 | Cenário externo sem bloqueios | Cenário sem bloqueio ao zênite e sem bloqueios relevantes em todo horizonte de observação (considerou-se, por padrão, que a soma dos bloqueios não deve totalizar >25% do horizonte de observação). Cenário tido como ideal para recepção de sinais GNSS. |
| 2 | Cenário externo com bloqueios relevantes | Ambientes sem bloqueio ao zênite, mas com bloqueios relevantes (i.e. >25%) que possam interferir sobre os satélites recebidos. |
| 3 | Cenário interno próximos de aberturas | Ambientes com bloqueios ao zênite e com o receptor posicionado em região de possível visada direta com um ou mais satélites (p.ex. devido a aberturas em geral - janelas, portas, claraboias, etc.). |
| 4 | Cenário interno padrão | Ambientes com bloqueios ao zênite e com receptor sem visadas diretas a possíveis satélites. |

Para cada tipo de cenário proposto foram selecionados 10 diferentes lugares com características visualmente bem conhecidas e de fácil distinção visual, totalizando 40 cenários diferentes de testes. Em cada lugar foram realizadas medições durante 1 hora, totalizando 3600 épocas de medições. A Tabela III apresenta os valores consolidados de média e desvio-padrão para as variáveis de média e soma de $C/N_0$ PDOP, HDOP e quantidade de satélites para um conjunto de medições em um determinado cenário. Sendo obrigatória uma medição por segundo, a coluna "Épocas" marca fora dos parênteses a quantidade de épocas para as quais se obtiveram medições e entre parênteses as épocas com medições ausentes (isto é, que não tenham apresentado ao menos um satélite com $C/N_0 > 0$).

TABELA III
RESULTADOS DOS TESTES ESTÁTICOS UTILIZANDO RECEPTORES GPS

| Cenário externo aberto | | | | | | |
|---|---|---|---|---|---|---|
| | Média $C/N_0$ | Soma $C/N_0$ | PDOP | HDOP | Satélites | Épocas |
| Local A-1 | 38.84 ± 1.00 | 528.60 ± 68.52 | 1.91 ± 0.31 | 0.87 ± 0.04 | 13.61 ± 1.72 | 3527(73) |
| Local A-2 | 38.14 ± 1.24 | 485.02 ± 88.39 | 1.99 ± 0.19 | 0.83 ± 0.04 | 12.73 ± 2.35 | 3598 (2) |
| Local A-3 | 37.61 ± 1.25 | 488.97 ±133.95 | 1.76 ± 0.21 | 0.85 ± 0.05 | 13.02 ± 3.64 | 3272(328) |
| Local A-4 | 32.67 ± 1.94 | 528.64 ±169.65 | 1.65 ± 0.21 | 0.91 ± 0.10 | 16.15 ± 4.95 | 3486(124) |
| Local A-5 | 36.33 ± 1.40 | 474.55 ± 99.65 | 1.85 ± 0.21 | 0.89 ± 0.06 | 13.09 ± 2.83 | 3419 (181) |
| Local A-6 | 34.76 ± 1.50 | 435.12 ± 66.42 | 1.68 ± 0.23 | 0.90 ± 0.11 | 12.51 ± 1.81 | 3558 (42) |
| Local A-7 | 39.94 ± 1.23 | 523.16 ±192.32 | 1.86 ± 0.29 | 0.88 ± 0.07 | 13.07 ± 4.72 | 3065 (535) |
| Local A-8 | 39.83 ± 1.09 | 566.65 ± 39.60 | 1.78 ± 0.08 | 0.86 ± 0.07 | 14.23 ± 1.02 | 3578(22) |
| Local A-9 | 38.03 ± 1.14 | 507.87 ±73.41 | 2.06 ± 0.21 | 0.83 ± 0.03 | 13.36 ± 1.96 | 3552 (48) |
| Local A-10 | 38.86 ± 3.12 | 571.01 ±123.50 | 1.75 ± 0.23 | 0.91 ± 0.11 | 14.68 ± 3.12 | 3404 (196) |
| Cenário externo com bloqueios parciais | | | | | | |
| | Média $C/N_0$ | Soma $C/N_0$ | PDOP | HDOP | Satélites | Medições |
| Local B-1 | 27.50 ± 1.97 | 291.34 ±29.79 | 3.09 ± 2.30 | 1.93 ± 2.12 | 10.64 ± 1.31 | 3599 (1) |
| Local B-2 | 27.10 ± 2.48 | 233.80 ± 25.38 | 3.33 ± 1.37 | 1.91 ± 1.24 | 8.70 ± 1.24 | 3600(0) |
| Local B-3 | 28.83 ± 2.34 | 248.28 ± 26.30 | 3.67 ± 1.21 | 2.50 ± 1.08 | 8.69 ±1.21 | 3600 (0) |
| Local B-4 | 25.76 ± 2.09 | 260.65 ± 27.24 | 2.22 ± 0.59 | 1.21 ± 0.25 | 10.14 ± 1.03 | 3598(2) |
| Local B-5 | 26.98 ± 2.65 | 281.10± 29.48 | 3.81± 6.53 | 2.57 ± 5.81 | 10.50 ± 1.35 | 3600(0) |
| Local B-6 | 29.82 ± 1.77 | 287.94 ± 22.08 | 2.07 ± 0.19 | 0.99 ± 0.10 | 9.67 ± 0.73 | 3600(0) |
| Local B-7 | 29.40 ± 1.72 | 319.90 ± 17.18 | 2.02 ± 0.28 | 0.96 ± 0.10 | 10.91 ± 0.82 | 3600 (0) |
| Local B-8 | 29.05 ± 2.37 | 263.98 ± 36.02 | 3.38 ± 0.54 | 1.88 ± 0.38 | 9.10 ± 1.17 | 3600 (0) |
| Local B-9 | 29.83 ± 2.41 | 254.56 ± 23.76 | 3.74 ± 1.99 | 2.77 ± 1.80 | 8.58 ± 1.04 | 3598 (2) |
| Local B-10 | 27.24 ± 2.09 | 264.66 ± 22.47 | 2.60 ± 0.33 | 1.32 ± 0.22 | 9.77 ± 1.13 | 3599(1) |
| Cenário interno próximo de aberturas | | | | | | |
| | Média $C/N_0$ | Soma $C/N_0$ | PDOP | HDOP | Satélites | Medições |
| Local C-1 | 20.64 ± 2.51 | 123.48 ± 40.75 | 22.38±38.18 | 21.11±38.78 | 6.08 ± 2.09 | 3521(79) |
| Local C-2 | 21.55 ± 2.05 | 172.95 ± 19.91 | 2.89±0.81 | 1.42±0.44 | 8.05 ± 0.85 | 3600 (0) |
| Local C-3 | 21.95 ± 3.30 | 118.43 ± 19.26 | 16.79±26.99 | 15.20±27.52 | 5.51 ± 1.16 | 3599 (1) |
| Local C-4 | 26.78 ± 2.63 | 122.69 ± 17.26 | 14.81±24.69 | 13.70±25.12 | 4.63 ± 0.85 | 3600 (0) |
| Local C-5 | 21.76 ± 2.44 | 102.60 ± 28.37 | 51.59±47.25 | 51.51± 49.34 | 4.73 ± 1.30 | 3600(0) |
| Local C-6 | 25.69 ± 2.13 | 159.13 ± 17.66 | 4.80±4.72 | 3.93±4.66 | 6.23 ± 0.86 | 3592 (8) |
| Local C-7 | 16.93 ± 1.71 | 157.17 ± 22.45 | 6.70±7.67 | 5.67±3.61 | 6.27 ± 1.11 | 3600 (0) |
| Local C-8 | 25.45 ± 3.31 | 149.53 ± 30.73 | 13.90±25.26 | 12.12±25.78 | 6.02 ± 1.49 | 3600(0) |
| Local C-9 | 22.18 ± 2.22 | 140.96 ± 18.66 | 5.19±9.42 | 3.91±9.32 | 6.39 ± 0.92 | 3600(0) |
| Local C-10 | 23.91 ± 2.29 | 174.32 ± 25.48 | 8.96±19.41 | 7.58±19.64 | 7.35 ± 1.24 | 3600(0) |
| **Cenário interno** | | | | | | |

|  | Média C/N$_0$ | Soma C/N$_0$ | PDOP | HDOP | Satélites | Medições |
|---|---|---|---|---|---|---|
| Local D-1 | 21.87 ±3.15 | 45.66 ± 27.12 | 99.99 ± 0.0 | 99.99 ± 0.0 | 2.07 ± 1.17 | 2965 (635) |
| Local D-2 | 20.84 ± 6.47 | 21.77 ± 7.44 | 99.99 ± 0.0 | 99.99 ± 0.0 | 1.05 ± 0.22 | 435 (3165) |
| Local D-3 | 21.39 ± 3.01 | 84.30 ± 27.47 | 54.27±46.48 | 52.70±47.65 | 4.04 ± 1.41 | 3600 (0) |
| Local D-4 | 23.73 ± 2.69 | 65.70 ± 36.13 | 99.99 ± 0.0 | 99.99 ± 0.0 | 2.79 ± 1.54 | 3377(223) |
| Local D-5 | 22.54 ± 2.64 | 73.93 ± 48.63 | 99.99 ± 0.0 | 99.99 ± 0.0 | 3.25 ± 2.04 | 3354 (246) |
| Local D-6 | 19.95 ± 4.23 | 35.67 ± 18.43 | 99.99 ± 0.0 | 99.99 ± 0.0 | 1.80 ± 0.91 | 2850 (750) |
| Local D-7 | 19.85 ± 5.69 | 70.79 ± 93.16 | 80.12±39.10 | 79.91±39.51 | 2.93 ± 3.09 | 2288 (712) |
| Local D-8 | 19.78 ± 4.53 | 27.37± 22.01 | 99.99 ± 0.0 | 99.99 ± 0.0 | 1.37 ± 0.76 | 2058 (1542) |
| Local D-9 | 19.41 ± 5.89 | 21.75 ± 9.40 | 99.99 ± 0.0 | 99.99 ± 0.0 | 1.12 ± 0.37 | 612 (2988) |
| Local D-10 | 18.82 ± 4.38 | 25.88 ± 13.61 | 99.99 ± 0.0 | 99.99 ± 0.0 | 1.36 ± 0.63 | 1913 (1687) |

Nos testes realizados, a média do valor de C/N$_0$ mostra-se adequada para caracterizar cenários quando possui valores extremos. Isto ocorre, em especial, em cenários abertos sem bloqueios (média de C/N$_0$ > 30 dB-Hz) e em cenários internos em geral (média de C/N$_0$ ≤ 25 dB-Hz). Em cenários internos com aberturas e externos com bloqueios (que apresentaram média no intervalo entre 20 e 30), o uso da média dos valores de C/N$_0$ não permitiu a distinção entre estes cenários de forma adequada.

Neste sentido, o uso do valor de soma de C/N$_0$ apresentou-se como uma métrica mais relevante para caracterizar tipos de cenários, sendo que, dentro dos testes realizados, pode ser utilizada como única métrica para caracterização de cenários. Nos resultados obtidos, somas de C/N$_0$ > 200 permitem caracterizar um local como um cenário externo e abaixo deste valor como um cenário interno, o que concorda com [7]. Enquanto em [7] distingue-se apenas os tipos mais gerais de cenário, verifica-se nos testes realizados nesta pesquisa que o uso da soma também permite caracterizar os limites entre cenários: valores de soma de C/N$_0$ >350 podem ser associados a ambientes externos abertos, enquanto valores entre 200 a 350 podem ser associados a regiões próximas de possíveis bloqueios. De forma contrária, valores entre 100 e 200 permitem caracterizar ambientes internos próximos de aberturas, enquanto valores menores podem ser associados a ambientes internos mais profundos.

Em termos dos valores de DOP, verifica-se que a utilização de PDOP e HDOP nos resultados apresentados podem ser utilizados para caracterizar ambientes internos e externos, considerando que regiões mais internas apresentam piores valores de geometria, enquanto regiões abertas possuem valores melhores de DOP (< 7). No entanto, como esta associação nem sempre pode ser confirmada devido ao comportamento de outras variáveis como, por exemplo, a relação entre as posições pontuais dos satélites e bloqueios diversos do ambiente em um determinado instante (que podem oferecer DOP ruim independentemente do tipo de cenário). Devido a estas limitações, o valor de DOP pode ser utilizado como uma métrica auxiliar para o processo de detecção, mas não se recomenda seu uso de forma exclusiva.

Nos testes realizados, a quantidade de satélites confirma o anteriormente citado que, em cenários com menores quantidades de bloqueios, a quantidade de satélites tende a ser maior. No entanto, como também citado previamente, isto pode ser de difícil mensuração no sentido que os sinais dos satélites podem ser obtidos de forma refletida ou de maneira fraca em ambientes internos e nos ambientes externos bloqueios podem limitar consideravelmente a oferta de satélites disponível [5].

A quantidade de épocas com medições não é um dado contido em uma mensagem NMEA, mas pode ser uma métrica auxiliar para detecção de tipos cenário: se um receptor está em um cenário com grande número de bloqueios, possivelmente o receptor não está obtendo dados adequados e, possivelmente, está em um ambiente interno longe de possíveis aberturas. No entanto, como a ausência de medições pode também decorrer por diversos motivos operacionais, esta métrica deve ser utilizada de forma criteriosa e auxiliar a outras métricas mais consistentes.

Considerando os testes e os dados obtidos é possível parametrizar os valores associados a cada tipo de cenário. Conforme visto previamente, o parâmetro de soma de C/N$_0$, de forma exclusiva, pode ser utilizado como um recurso para detecção de cenários externos e internos (Tabela IV).

TABELA IV
PARÂMETROS PARA DETECÇÃO DE CENÁRIOS INTERNOS E EXTERNOS UTILIZANDO SOMA DOS SINAIS DE C/N$_0$

| Cenários | Parâmetros |
|---|---|
| Cenário externo aberto | Soma de C/N$_0$ ≥ 350 |
| Cenário externo com bloqueios | Soma de C/N$_0$ ≥ 200 e < 350 |
| Cenários internos próximos de aberturas | Soma de C/N$_0$ ≥ 100 e < 200 |
| Cenários internos | Soma de C/N$_0$ ≥ 0 e < 100 |

Considerando outras métricas, os seguintes valores podem ser utilizados para tornar a caracterização mais consistente (Tabela V).

TABELA V
PARÂMETROS PARA DETECÇÃO DE CENÁRIOS INTERNOS E EXTERNOS UTILIZANDO MÉTRICAS COMBINADAS

| Cenários | Parâmetros | | |
|---|---|---|---|
| Cenário externo aberto | Soma de C/N$_0$ ≥ 350 e Média de C/N$_0$ ≥ 30 | | (PDOP e HDOP) ≤ 7 |
| Cenário externo com bloqueios | Soma de C/N$_0$ ≥ 200 e < 350 | Média de C/N$_0$ ≥ 20 e <30 | |

| Cenários internos próximos de aberturas | Soma de $C/N_0 \geq 100$ e $< 200$ | | (PDOP e HDOP) >7 |
|---|---|---|---|
| Cenários internos | (Soma de $C/N_0 < 100$ e Média de $C/N_0 < 25$) ou Nr. Satélites <4 | | |

*B. Validação dos parâmetros obtidos*

A segunda fase de testes envolveu a validação dos parâmetros anteriormente obtidos. Nesta fase, foi realizado um conjunto de medições em diferentes lugares com tipos bem conhecidos. Sobre estas medições, aplicaram-se os seguintes testes de validação: (1) - utilizando como parâmetro apenas a soma de $C/N_0$ e (2) - utilizando os parâmetros combinados (conforme a Tabela IV). Com a aplicação dos parâmetros, obteve-se para cada época de medição o tipo ao qual o cenário deveria pertencer. Este resultado foi comparado com o tipo conhecido do local em questão e a Tabela VI apresenta a concordância dos resultados.

TABELA VI
RESULTADOS DOS TESTES DE VALIDAÇÃO

| Teste de Validação | Conjunto de épocas | Acertos (%) | Erros (%) |
|---|---|---|---|
| 1 | 82323 | 80765 (98.11%) | 1558 (1.89%) |
| 2 | 82323 | 73650 (89.46%) | 8673 (10.54%) |

Em ambas validações, os principais erros de caracterização ocorreram, em especial, nos limites entre os cenários internos e externos. Os limites entre estes cenários são, em geral, frágeis e de difícil caracterização direta. Para a maior parte das atividades de posicionamento e navegação, no entanto, a caracterização perfeita pode não ser necessária já que se pressupõe que um receptor em cenários internos próximos de aberturas e em cenários externos com bloqueios relevantes está em condições que necessitam de outros recursos para garantir a geração adequada da solução de posicionamento.

Nos testes de validação realizados, a utilização da soma de $C/N_0$ apresentou, isoladamente, resultados melhores do que a utilização de combinação de parâmetros. Mesmo apresentando maior número de erros, a utilização de múltiplos parâmetros combinados logicamente torna mais consistente, não-ambígua e exigente a determinação de um dado cenário e pode ser uma abordagem mais recomendada para atividades críticas de caracterização.

O uso isolado de soma de $C/N_0$, apesar de aparentemente simples, apresenta as características de outras métricas: por exemplo, para obter uma soma superior a 350 dB-Hz, em geral, necessita-se de um mínimo de 9-10 satélites com valores de $C/N_0$ superiores a 35 dB-Hz. Desta forma, é possível determinar que se o receptor consegue medir dez satélites com estas características, os requisitos de média de $C/N_0$, de quantidade de satélites e de geometria são atendidos. No entanto, em regiões de bloqueios (internas e externas) a combinação de sinais diretos e indiretos dos satélites podem distorcer a média de valor de $C/N_0$, além dos valores de número de satélites e DOP não oferecem padrões convenientes para caracterizar o tipo de cenário. Mesmo sob estas condições, a soma de $C/N_0$ apresenta características adequadas para distinguir um cenário específico.

IV. CONSIDERAÇÕES FINAIS

Este artigo apresentou a utilização de dados extraídos de mensagens NMEA para caracterização de ambientes internos e externos com o objetivo de fornecer mecanismos de contextualização para sistemas de POS/NAV. Foram realizados um conjunto de testes e diversas métricas foram avaliadas para buscar detectar o tipo de cenário de operação de receptor GPS. Nos testes realizados, a soma de dados de $C/N_0$ para uma determinada época apresentou-se como a opção mais consistente, ainda que outras métricas como média de $C/N_0$, dados de DOP, quantidade de satélites e medições possam também serem utilizadas para suportar esta atividade. A técnica, assim, mostrou-se adequada para tratamento deste tipo de caracterização, oferecendo menores recursos do que técnicas mais elaboradas como aquelas apresentadas em [6].

Possibilidades de trabalhos futuros decorrentes desta pesquisa podem envolver a avaliação de uso de receptores multiconstelação para determinar os parâmetros relacionados a cada cenário, assim como análises de outros dados como observáveis GNSS (pseudodistância, análise da onda portadora, etc.). Adicionalmente, outras possibilidades de trabalhos futuros podem avaliar os parâmetros obtidos em atividades cinemáticas, assim como utilização integrada de outras tecnologias como sistemas inerciais (para determinar se o receptor está parado ou em movimento) e sinais de outras tecnologias de rádio (e.g. Wi-Fi, sinais de oportunidade).

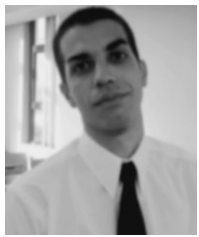
**Rodrigo de Sousa Pissardini** é bacharel em Ciência da Computação pelo Centro Universitário São Camilo (2009) e Mestre em Engenharia de Transportes pela Escola Politécnica da Universidade de São Paulo (2014). Desde 2015 é doutorando em Engenharia de Transportes pela mesma instituição. Possui experiência nas áreas de gerenciamento de projetos de tecnologia da informação, mineração de dados e inteligência artificial. Suas pesquisas atuais concentram-se em veículos autônomos, posicionamento GNSS e posicionamento em cenários urbanos.

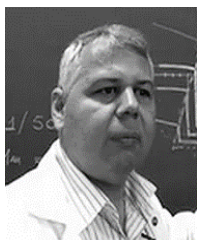
**Edvaldo Simões da Fonseca Junior** possui graduação em Engenharia Cartográfica pela Universidade do Estado do Rio de Janeiro (1985), mestrado (1996), doutorado (2002) e Livre Docência (2017) em posicionamento por satélites junto ao programa de pós-graduação em Engenharia de Transportes da Escola Politécnica da Universidade de São Paulo. Atualmente é professor associado do Departamento de Engenharia de Transportes da Escola Politécnica da Universidade de São Paulo. Tem experiência em Geociências com ênfase em Geodesia, atuando principalmente nos temas: GNSS, posicionamento por satélites, monitoramento de estruturas com instrumentos geodésicos, navegação autônoma e controle de veículos.